# Quantum oscillations in a dipolar excitonic insulator

Phuong X. Nguyen[1,2], Raghav Chaturvedi[1], Bo Zou[3], Kenji Watanabe[4], Takashi Taniguchi[4], Allan H. MacDonald[3], Kin Fai Mak[1,2,5,6*], Jie Shan[1,2,5,6*]

[1]School of Applied and Engineering Physics, Cornell University, Ithaca, NY, USA
[2]Kavli Institute at Cornell for Nanoscale Science, Ithaca, NY, USA
[3]Department of Physics, University of Texas at Austin, Austin, TX, USA
[4]National Institute for Materials Science, Tsukuba, Japan
[5]Laboratory of Atomic and Solid State Physics, Cornell University, Ithaca, NY, USA
[6]Max Planck Institute for the Structure and Dynamics of Matter, Hamburg, Germany.

[*]Email: kinfai.mak@cornell.edu; jie.shan@cornell.edu

These authors contributed equally: Phuong X. Nguyen, Raghav Chaturvedi

**Quantum oscillations in magnetization or resistivity are a defining feature of metals subject to an external magnetic field. The phenomenon is generally not expected in insulators without a Fermi surface. The observations of quantum oscillations in Kondo insulating materials[1-5] have provided a rare counterexample and attracted much theoretical interest[6-12]. However, the magnetic oscillations in correlated insulators remain poorly understood. Here we report the observations of resistivity quantum oscillations in an excitonic insulator realized in Coulomb-coupled electron-hole double layers[13-21] with gate-tunability that allows the phenomenon to be explored in a more controllable fashion than in bulk materials. When the cyclotron energy of the electrons or holes is tuned to be comparable to or larger than the exciton binding energy, recurring transitions between excitonic insulators and electron-hole decoupled quantum Hall states are observed. Compressibility measurements show an oscillatory exciton binding energy as a function of magnetic field and electron-hole pair density. Coulomb drag measurements further reveal the formation of excitons with finite angular momentum. Our results are qualitatively captured by mean-field theory calculations[22,23]. The study demonstrates a new platform for studying quantum oscillations in correlated insulators.**

## Main

The observation of quantum oscillations in magnetization or resistivity in insulators[1-5,24-28] has intrigued the condensed matter physics community because it is long believed that the phenomenon could occur only in metals with a well-defined Fermi surface. Many theoretical proposals have been put forth to explain the observation, including magnetic field-induced gap oscillations in both single-particle and correlated insulators[6,8,9,12,22,23,29-33], competing many body ground states as a function of magnetic field[22,23,31,32] and an exotic fractionalization scenario with neutral fermions and gauge fields[10,11]. To date, the mechanism responsible for oscillations in different material systems remains under debate, especially so for strongly correlated systems. Limits on the ability to control the properties of many of the bulk correlated materials has hindered the development of a cohesive understanding. The emergence of highly tunable correlated insulating states in van der

Waals' heterostructures provides an opportunity to study this problem from a fresh perspective.

In this work, we report the observations of quantum oscillations in the resistivity, exciton binding energy and exciton currents in a dipolar excitonic insulator (EI). This is achieved through a suite of magneto-transport, capacitance and drag counterflow measurements performed on Coulomb-coupled MoSe$_2$/WSe$_2$ electron-hole double layers. The quantum oscillations originate from recurring transitions between competing EI and layer-decoupled quantum Hall (QH) states as a function of magnetic field and electron-hole pair density[22,23,32]. Compared to earlier reports on the observations of quantum oscillations in the EI candidate InAs/GaSb quantum well[24-26], the ability to electrically tune and address the electron and hole layers separately has allowed us to establish the emergence of quantum oscillations in an EI.

Figure 1a shows a schematic cross-section of a dual-gated electron-hole double layer device. The double layer consists of a natural bilayer MoSe$_2$ and a monolayer WSe$_2$ separated by a thin hexagonal boron nitride (hBN) barrier (5-6 layers thick). The MoSe$_2$/WSe$_2$ heterobilayer has a type-II band alignment[15-19,21] (Fig. 1b) with the conduction (valence) band minimum (maximum) residing in the Mo-layer (W-layer). The two gates allow independent control of the net electron doping density ($n$) in the double layer and the electric field ($E$) perpendicular to the sample plane. In our experiment, we apply a constant electric field to reduce the heterobilayer energy gap ($E_g$) to $E_g \approx 0.625$ eV from its zero-field value $E_g \approx 1.6$ eV (Ref. [16,18,21]). We also apply a bias voltage ($V_b$) between the Mo- and W-layers to separate the electron and hole Fermi levels. When the effective charge gap of the double layer ($E_g - eV_b$) becomes smaller than the interlayer exciton binding energy ($E_b$), the double layer is spontaneously populated by interlayer excitons (with an exciton chemical potential $eV_b - E_g$), forming an EI with a net electric polarization[34-37] (i.e. a dipolar EI). ($e$ is the electron charge.) The electrons in the Mo-layer and holes in the W-layer are coupled only through interlayer Coulomb interactions; the single-particle interlayer tunneling is negligible as verified by negligible interlayer tunneling current measured in our devices (Extended Data Fig. 3).

Landau levels (LLs) emerge in the electron and hole layers under a perpendicular magnetic field, $B$, (Fig. 1c,d). When the excitons are dissociated into an electron-hole plasma at pair densities ($n_p$) higher than the exciton Mott density ($n_M$)[16,18,34,35], layer-decoupled QH states for the electrons and holes are expected. On the other hand, the EI is expected to dominate at low pair densities when $E_b$ is large compared to the electron/hole cyclotron energy ($E_c$). (The electron and hole cyclotron energies are comparable.) Near $n_p \approx n_M$, where $E_c$ is comparable to $E_b$, it was suggested in Ref. [22,23] that recurring transitions between the EI and layer-decoupled QH states could emerge as a function of $n_p$ and $B$. Unlike the interlayer-coherent exciton condensates in half-filled LLs in electron-electron or hole-hole Coulomb-coupled bilayers[38-42], in which the condensates carry chiral edge states (i.e. they are quantum Hall ferromagnets) and are unstable in the $B = 0$ T limit[22], the EI in electron-hole double layers is stable in the zero-field limit and does not support chiral edge states because interlayer coherence is expected to gap out the counter-propagating edge states in the electron and hole layers[22,23].

We probe the physics under finite $B$ by patterning the double layer into a Hall bar device with independent electrical contacts to the Mo- and W-layers. Both the four-terminal longitudinal ($R_{xx}$) and Hall ($R_{xy}$) resistances can be measured in the W-layer but only the two-terminal resistance can be accessed in the Mo-layer due to the less reliable electron contacts formed via Bi evaporation. In our experiment, we measure $R_{xx}$ and $R_{xy}$ in the W-layer while keeping the Mo-layer in open-circuit (Fig. 1e). Although only one of the layers is measured, $R_{xx}$ and $R_{xy}$ are highly sensitive to interlayer Coulomb interactions, as shown by a recent study[21]. We also supplement this measurement by studies in the drag counterflow geometry (Fig. 1f), which has been shown to directly probe interlayer Coulomb interactions[18,19,40], and by penetration capacitance measurements[16], which can access the thermodynamic gap size of the double layer. See Methods for details on device fabrication and electrical transport and capacitance measurements.

**Resistivity quantum oscillations in EIs**

Figure 1g shows $R_{xx}$ of the W-layer (Mo-layer in open circuit) as a function of $V_g \equiv \frac{V_{tg}+V_{bg}}{2}$ and $V_b$ at $B = 0$ T (for device 1). Unless otherwise specified, the sample temperature is at $T = 1.5$ K. Here $V_{tg}$ and $V_{bg}$ denote the top and bottom gate voltages, respectively. For a device with nearly symmetric top and bottom gates, the symmetric part $V_g = \frac{V_{tg}+V_{bg}}{2}$ tunes the net doping density $n$ in the double layer, and the bias voltage $V_b$ largely controls the electron-hole pair density $n_p$ (Ref. [16]). The different regions in the electrostatics phase diagram are labeled as *ii*, *in*, *pi* and *pn* with *i*, *p* and *n* denoting an intrinsic (or charge neutral), hole-doped and electron-doped layer, respectively; the first and second entries represent the doping state in the W- and Mo-layer, respectively. As expected, $R_{xx}$ diverges at low temperatures in the *ii* and *in* regions (grey-shaded); the W-layer is turned on electrically only in the *pi* and *pn* regions (the *pi-ii* boundary is vertical because the W-layer is grounded). $R_{xx}$ also diverges in the EI region (the triangular tip of the *ii* region that protrudes into the *pn* region) because the bound electron-hole pairs and the open-circuit Mo-layer forbid the flow of a hole current in the W-layer[21]. The metal-insulator transition at the tip of the EI triangle corresponds to an exciton Mott transition[16,18], in which the interlayer excitons are dissociated into an electron-hole plasma; the unbound holes in the plasma give a metallic W-layer. The results are fully consistent with earlier studies[16,18,21].

Next, we examine the $R_{xx}$ and $\sigma_{xy}$ maps versus $V_g$ and $V_b$ at $B = 12$ T (Fig. 2a,b). ($\sigma_{xy}$ is the Hall conductivity.) Maps at other magnetic fields up to 31.5 T are shown in Extended Data Fig. 4. LLs are observed in both the *pi* and *pn* regions of the phase diagram; every fully filled LL corresponds to a $R_{xx}$ dip and a nearly quantized $\sigma_{xy}$ at $\frac{v_h e^2}{h}$ (see line cuts in Fig. 2d). (Here $v_h$ and $h$ denote the LL filling factor in the W-layer and the Planck's constant, respectively.) The LLs form vertical stripes in the *pi* region because only the W-layer is doped so that its doping density is independent of $V_b$. Those in the *pn* region form diagonal stripes because the LLs in the Mo-layer are also populated and $v_e$ changes with both $V_g$ and $V_b$ ($\nu_e$ is the LL filling factor in the Mo-layer). The LLs in the Mo-layer are nearly invisible because only $R_{xx}$ and $\sigma_{xy}$ of the W-layer are measured; their presence can be observed in drag counterflow measurements (see Fig. 5 below) and in capacitance measurements (Extended Data Fig. 5).

The most interesting features in the $R_{xx}$ and $\sigma_{xy}$ maps are located near the EI triangle of the phase diagram, where the electron and hole densities are nearly the same. Near the bottom of the triangle ($0.60 \lesssim V_b \lesssim 0.64$ V), where the exciton binding energy is large, the diagonal LL stripes are in general interrupted by the EI at $\nu_e \approx \nu_h$, where $R_{xx}$ diverges and $\sigma_{xy}$ is zero, i.e. a transition from QH states to an EI. Only at the highest magnetic fields the LL stripes can penetrate through the bottom of the EI triangle (Extended Data Fig. 4). In contrast, the LL stripes are uninterrupted above the tip of the triangle ($V_b \gtrsim 0.66$ V), where the system is an electron-hole plasma without exciton binding ($E_b = 0$), i.e. only QH states remain. Near the tip of the triangle ($0.64 \lesssim V_b \lesssim 0.66$ V), a fully-filled LL (one with a $R_{xx}$ dip and quantized $\sigma_{xy}$) can nearly penetrate through the EI phase without interruption even under moderate magnetic fields; on the other hand, a half-filled LL (one with a $R_{xx}$ peak) transitions to an EI with much higher $R_{xx}$ and strongly suppressed $\sigma_{xy}$.

Figure 2c-e show representative line cuts along the solid lines labelled in the maps. Only a series of QH states is observed away from charge neutrality at $\nu_e - \nu_h = 3$ (Fig. 2d). At charge neutrality ($\nu_e = \nu_h$), the transition from the $\nu_h = 6$ ($V_b \approx 0.662$ V) to $\nu_h = 5$ ($V_b \approx 0.645$ V) QH state is interrupted by an EI with enhanced $R_{xx}$ and strongly suppressed $\sigma_{xy}$ at $V_b \approx 0.655$ V; only an EI with diverging $R_{xx}$ and zero $\sigma_{xy}$ is stable for $V_b \lesssim 0.64$ V (Fig. 2c). An EI at $V_g \approx -0.01$ V interrupting the transition from the $\nu_h = 5$ ($V_g \approx -0.05$ V) to $\nu_h = 4$ ($V_g \approx 0.03$ V) QH state is also illustrated by a line cut at constant $V_b \approx 0.64$ V (Fig. 2e).

We now examine the magnetic field dependence at charge neutrality ($\nu_e = \nu_h$) in Fig. 3a, i.e. along the yellow solid line in Fig. 2a, up to $B = 31.5$ T. With increasing field, the EI region expands; the exciton Mott transition boundary (black dashed line), below (above) which $R_{xx}$ is insulating (metallic), moves to higher $V_b$ in a hyperbolic fashion due to exciton diamagnetic shift[43]. LLs emerge in the electron-hole plasma region at $B \gtrsim 3$ T. The LL dispersion with $V_b$ is slightly nonlinear because of the strong $V_b$-dependence for the interlayer capacitance (between the Mo- and W-layer, see Ref. [16]). At high fields, the EI phase with large $R_{xx}$ is interrupted by LLs labeled by $\nu_e = \nu_h = 1, 2, 3$ and $4$ (the $R_{xx}$ dips) with decreasing onset fields for increasing $\nu_e$ or $\nu_h$. The horizontal $R_{xx}$ stripes in the EI phase are caused by the LLs in the graphite gates and by the capacitive coupling between the W-layer and the top graphite gate (see Ref. [27,28]); they are independent of $V_b$ because the W-layer is grounded; they are distinct from the LLs emerging from the band edge of the W-layer.

We further study the temperature dependence at constant $B = 12$ T as a function of $V_b$ (Fig. 4a). Representative line cuts at different temperatures are shown in Fig. 4b. Recurring transitions between EI and QH states are observed as $V_b$ increases; $R_{xx}$ increases (decreases) with decreasing $T$ for the EI (QH state) and is nearly $T$-independent at the boundary separating the two states. We also extract the thermodynamic gap size versus $V_b$ for two different devices (device 1 and 2) using penetration capacitance measurements (Methods and Extended Data Fig. 5). Both devices show similar results. The gap size oscillates as a function of $V_b$ on top of a monotonic decreasing background (Fig. 4c); it reaches a local minimum at the QH states.

The results above demonstrate magnetic field and pair density (or $V_b$) tuned transitions between EIs and QH states, which give rise to the observed quantum oscillations in the EI region of the phase diagram. The interlayer exciton binding energy $E_b$ decreases continuously with the pair density $n_p$ (or $V_b$) due to screening[16,18,44] while the cyclotron energy $E_c$ in each layer increases with $B$. When $E_b$ becomes comparable to or smaller than $E_c$, the interlayer-coherent EI is expected to transition to layer-decoupled QH states when both the Mo- and W-layers are in fully filled LLs (i.e. $\nu_e = \nu_h =$ integers)[22,23]. The EI phase returns when the LLs are partially filled until $n_p$ exceeds the Mott density $n_M$, beyond which excitons are no longer stable and only layer-decoupled LLs are present in the electron-hole plasma.

We qualitatively capture the experimental observations by performing mean-field calculations for the ground state energy density of the electron-hole double layer[22] (Methods). The theoretical phase diagram (Fig. 3c) is in good qualitative agreement with the experimental phase diagram (Fig. 3a). In particular, a fan of layer-decoupled QH states at integer $\nu_e = \nu_h$ emerges at high magnetic fields; the onset fields for the QH states decrease with increasing $\nu_e$ or $\nu_h$. However, the magnetic field scale is substantially higher than that in experiment mainly because the pair-density-dependent exciton binding energy due to screening effects and the associated exciton Mott transition have not been properly accounted for in the mean-field theory. Future studies taking these effects into consideration are required to quantitatively describe the experimental results.

**Quantum oscillations in Coulomb drag**

Finally, we perform drag counterflow measurements to further examine the EI to QH state transition. We drive an AC bias current ($I_{drive}$) in the Mo-layer and measure the drag current ($I_{drag}$) in the W-layer[18] (Fig. 1f); the interlayer tunneling current is negligible compared to $I_{drive}$ and $I_{drag}$ (Extended Data Fig. 3). The drag ratio $\frac{I_{drag}}{I_{drive}}$, as shown in Fig. 3b for device 3, provides a measure for the interlayer Coulomb coupling[18]. In particular, near perfect Coulomb drag ($\frac{I_{drag}}{I_{drive}} \approx 1$) is observed for the EI phase; near zero drag ($\frac{I_{drag}}{I_{drive}} \approx 0$) is observed for the layer-decoupled QH states (labeled by $\nu_e = \nu_h = 2, 3$ and 4) and the electron-hole plasma. The results are fully consistent with the $R_{xx}$ phase diagram in Fig. 3a.

We also examine the behavior of Coulomb drag in the presence of an electron/hole density imbalance. Figure 5a and 5b show $I_{drive}$ and $I_{drag}$ in the Mo- and W-layer, respectively, as a function of $V_g$ and $V_b$ for device 2 at $B = 12$ T. The Mo-layer is turned on electrically only in the *in* and *pn* regions. A large $I_{drag}$ in the W-layer with nearly the same magnitude as, but with opposite sign to $I_{drive}$ is observed in the triangular EI region. Moreover, quantum oscillations in $I_{drag}$ are observed near the boundaries of and outside the EI triangle. Representative line cuts for $I_{drive}$, $I_{drag}$ and $\frac{I_{drag}}{I_{drive}}$ along the $\nu_e = \nu_h$ and $\nu_e - \nu_h = -1$ lines (blue and yellow, respectively, in Fig. 5a) are shown in Fig. 5c. Near perfect Coulomb drag ($\frac{I_{drag}}{I_{drive}} \approx 1$) is observed at $\nu_e = \nu_h$ for $V_b \lesssim 0.55$ V, above which $\frac{I_{drag}}{I_{drive}}$ drops quickly and exhibits quantum oscillations. A large but imperfect drag ratio (up to about 0.5) is also observed at $\nu_e - \nu_h = -1$ near the onset of electron-hole injection; $\frac{I_{drag}}{I_{drive}}$ also drops quickly and shows quantum oscillations with increasing $V_b$.

We trace the trajectories for $\frac{I_{drag}}{I_{drive}} \approx 0$ and construct a phase diagram for the layer-decoupled QH states labelled by $\nu_e$ and $\nu_h$ in Fig. 5d. The regions $\nu_e - \nu_h = -1, 0, 1$ and 2 are shaded. The EI (with $\frac{I_{drag}}{I_{drive}} \approx 1$) dominates in the $\nu_e = \nu_h$ region except near the tip of the triangle, where $\nu_e = \nu_h = 4, 5$ and 6. The presence of substantial Coulomb drags (with ratio up to about 0.5) at the partially filled LLs (i.e. in between the solid lines in Fig. 5d) in the $\nu_e - \nu_h = -1, 1$ and 2 regions demonstrates the emergence of interlayer excitons carrying angular momentum quantum $-\hbar$, $\hbar$ and $2\hbar$, respectively[22]. The results suggest the tendency to stabilize EIs with finite angular momentum excitons, as suggested by mean-field calculations[22]. However, the imperfect Coulomb drag shows the presence of ionized electrons and holes in these regions potentially due to thermal and/or quantum fluctuations. Further studies under higher magnetic fields are required to fully establish the emergence of finite angular momentum EIs.

**Conclusions**
To conclude, we observe resistivity and Coulomb drag quantum oscillations in a dipolar EI. The oscillations originate from phase transitions between the competing EI and layer-decoupled QH states in Coulomb-coupled MoSe$_2$/WSe$_2$ electron-hole double layers[22,23]. The results demonstrate a highly tunable platform for studying quantum oscillations in correlated insulators. The observed evidence of excitons with finite angular momentum also opens the door for realizing the quantum Hall effect for excitons.

**Methods**
**Device design and fabrication**
We have detailed the device and contact geometry in Ref. [16,18,21]. The optical images of device 1 and 2 are shown in Extended Data Fig. 1a and 1b, respectively. The top and bottom graphite gates are outlined in dashed and solid red lines, respectively. The WSe$_2$ monolayer (green) and MoSe$_2$ bilayer (blue) are separated by a thin hBN barrier (1.5-2 nm) in the transport channel (shaded red) and by a thick hBN barrier (10 nm) in the exciton contact regions (shaded pink). (A natural bilayer is chosen because it does not crack as easily as a monolayer during the fabrication process. Electrons in the bilayer are layer-polarized under a high perpendicular electric field employed in this study.) To achieve high doping densities at the metal-semiconductor contacts and thus ohmic contacts at low temperatures, the Pt-WSe$_2$ contacts are gated only by the top gate, and likewise only by the bottom gate for the Bi-MoSe$_2$ contact. During measurements, we maintained a constant $\Delta = 3\,V$ for device 1 and $\Delta = 5.5\ V$ for device 2. Here $\Delta \equiv \frac{V_{bg}-V_{tg}}{2}$ is the antisymmetric part of $V_{tg}$ and $V_{bg}$; it is proportional to the electric field $E$.

We fabricated the devices using the layer-by-layer dry transfer method described in Ref. [16,18,21,45]. In short, all the individual layers were first mechanically exfoliated by Scotch tape onto 285nm SiO$_2$/Si substrates and screened by optical microscopes for appropriate thickness and geometry before stacking. We picked up each layer sequentially using a polymer stamp made of a thin layer of polycarbonate on a polypropylene-carbonate-coated polydimethylsiloxane block. The completed stack was then released onto pre-prepattern Pt electrodes (on SiO$_2$/Si substrates) to make contacts to the W-layer and to the gate electrodes. The polymer residual was removed by chloroform and isopropanol. We then etched the top graphite gate to define separate contacts for the W-layer using electron-beam lithography patterning (Nabity) and oxygen plasma reactive-ion etching (Oxford Plasmalab80Plus). Finally, we made contacts to the Mo-layer by another electron-beam lithography patterning followed by Bi evaporation in a thermal evaporator[46]. We have reproduced the main results in three difference devices. See Extended Data Fig. 6 for data from device 3.

**Electrical measurements**
We examined the transport properties of the electron-hole double layer using two measurement configurations: the open-circuit $R_{xx}$ and $R_{xy}$ measurements (Fig. 1e) and the drag counterflow measurement (Fig. 1f), which reflect the charge and exciton transport properties, respectively. Details of the measurements have been reported in Ref. [18,21]. Extended Data Fig. 2a and 2b show the circuit diagrams for the measurements. In the open-circuit measurement (device 1, Fig. 1-3), the Mo-layer is kept open-circuit, i.e. no electron and therefore no exciton current can flow, and four-terminal measurements of $R_{xx}$ and $R_{xy}$ were performed on the W-layer using standard lock-in techniques. Here $R_{xx}$ and $R_{xy}$ measure only the transport properties of the unbound holes in the W-layer[21]. There are in total eight contacts in the W-layer defined by the etched top gate (numbered 1-8). To obtain $R_{xx}$, we applied a small AC bias voltage 0.15 mV (RMS) between pin 2 and 6 at 11.33 Hz and measured simultaneously the bias current out of pin 6 and the longitudinal voltage drop between pin 1 and 5 at the same modulation frequency. To obtain $R_{xy}$, we applied a small AC bias voltage 0.15 mV (RMS) between pin 1 and 4 at 11.33 Hz and measured simultaneously the bias current out of pin 4 and the Hall voltage drop between pin 2 and 6 at the same modulation frequency. We also confirmed that other longitudinal and Hall measurement configurations give similar results.

In the drag counterflow measurements (device 2 and 3, Fig. 3 and 4), both layers are kept closed-circuit so that excitons can flow in the transport channel. We drove an AC bias current in the Mo-layer and measured the drag current in the W-layer (Fig. 1f). The drag current ratio $\frac{I_{drag}}{I_{drive}}$ reflects the exciton population fraction in the system[18]; in particular, $\frac{I_{drag}}{I_{drive}} = 1$ and 0 correspond to pure exciton and pure charge transport, respectively. We applied an AC bias voltage 10 mV (RMS) at 7.33 Hz to drive a current in the Mo-layer through a 1:1 voltage transformer. The voltage transformer was connected to a 10 kΩ potentiometer to distribute the AC voltage on the two ends of the Mo-layer; this step minimized the AC coupling between the Mo- and W-layers[41,42]. The DC interlayer bias voltage $V_b$ was applied to the middle of the potentiometer; this kept both contacts in the Mo-layer at the same DC potential while maintaining a 10mV AC voltage drop between them[18]. We measured $I_{drive}$ by monitoring the voltage drop across a 150 kΩ resistor connected in series with the Mo-layer and $I_{drag}$ by an ammeter connected in series with the W-layer.

We carried out all the electrical measurements in a closed-cycle He$^4$ cryostat (Oxford TeslatronPT) with temperature down to 1.5 K and magnetic field up to 12 T. All the currents and voltages were measured by lock-in amplifiers from Stanford Research Systems (SR830 and SR860). In voltage measurements, the signals were first sent to an Ithaco DL 1201 preamplifier with input impedance 100 MΩ before being measured by the lock-in amplifiers. The measurement results are largely independent of the excitation amplitude (3–20 mV) and frequency (7 Hz to 37 Hz).

**Capacitance measurements**
Details of the capacitance measurements have been reported by Ref. [16]. In short, we measured the penetration capacitance ($C_P$) of the dual-gated device 1 and 2 under $B = 12$ T to obtain the dependence of the thermodynamic gap size on $V_b$ (Fig. 4c). This was achieved by applying an AC bias voltage 5 mV (RMS) at 737 Hz to the top gate, and measuring the induced charge carrier density in the bottom gate by a GaAs high-electron-mobility-transistor (HEMT) through a low-temperature capacitance bridge[47]. The result, $C_P$ as a function of $V_g$ and $V_b$, is shown in Extended Data Fig. 5. The thermodynamic gap can be obtained by integrating the normalized $C_P$ with respect to $V_g$ at each $V_b$, i.e. gap ≈ $\int (C_P/C_{gg}) \, dV_g$, where $C_{gg}$ is the gate-to-gate geometrical capacitance[16].

**Mean-field calculations**
To obtain the phase diagram shown in Fig. 3c, we performed mean-field calculations using the self-consistent Hartree-Fock approximation described in Ref. [22], and assumed uniform carrier density distribution in our solutions. The single-particle part of the Hamiltonian consists of LLs from the two semiconductor layers; for simplicity, we assumed equal electron and hole cyclotron energies (a good approximation given the similar electron and hole mass ≈ $0.4m_0$). ($m_0$ is the free electron mass.) The LL energy in the conduction (valence) band increases (decreases) with the level index. The bias voltage $V_b$ controls the effective energy difference between the conduction band minimum and the valence band maximum. We considered the screening effects from the gate electrodes on both the intralayer and interlayer Coulomb interactions between charge carriers. In momentum ($\boldsymbol{q}$) space, the intralayer and interlayer Coulomb interactions are $V_A(\boldsymbol{q}) = \frac{2\pi e^2}{\epsilon q} \tanh(qD)$ and $V_E(\boldsymbol{q}) = V_A(\boldsymbol{q}) e^{-qd}$, respectively. The factor $\tanh(qD)$ comes from screening from the gate electrodes. Here $d = 1.3d_0$ and $D = 1.3D_0$ are the effective distance between the two semiconductors layers and that between the two gates, respectively, accounting for the anisotropic dielectric constant of hBN ($\epsilon \approx 5.2$); $d_0 = 1.8$ nm and $D_0 = 8$ nm are the true geometric distances in the device. We also increased the dielectric constant by 50% to account for the correlated screening effect. Finally, we evaluated the Hartree contribution to the interaction as the potential energy of the capacitor and the exchange integrals following Ref. [22].

**References**
1   Li, G. *et al.* Two-dimensional Fermi surfaces in Kondo insulator SmB6. *Science* **346**, 1208-1212 (2014).
2   Tan, B. S. *et al.* Unconventional Fermi surface in an insulating state. *Science* **349**, 287-290 (2015).
3   Hartstein, M. *et al.* Fermi surface in the absence of a Fermi liquid in the Kondo insulator SmB6. *Nature Physics* **14**, 166-172 (2018).
4   Liu, H. *et al.* Fermi surfaces in Kondo insulators. *Journal of Physics: Condensed Matter* **30**, 16LT01 (2018).
5   Xiang, Z. *et al.* Quantum oscillations of electrical resistivity in an insulator. *Science* **362**, 65-69 (2018).
6   Knolle, J. & Cooper, N. R. Quantum Oscillations without a Fermi Surface and the Anomalous de Haas--van Alphen Effect. *Physical Review Letters* **115**, 146401 (2015).
7   Erten, O., Ghaemi, P. & Coleman, P. Kondo Breakdown and Quantum Oscillations in SmB6. *Physical Review Letters* **116**, 046403 (2016).
8   Zhang, L., Song, X.-Y. & Wang, F. Quantum Oscillation in Narrow-Gap Topological Insulators. *Physical Review Letters* **116**, 046404 (2016).
9   Pal, H. K., Piéchon, F., Fuchs, J.-N., Goerbig, M. & Montambaux, G. Chemical potential asymmetry and quantum oscillations in insulators. *Physical Review B* **94**, 125140 (2016).
10  Sodemann, I., Chowdhury, D. & Senthil, T. Quantum oscillations in insulators with neutral Fermi surfaces. *Physical Review B* **97**, 045152 (2018).
11  Chowdhury, D., Sodemann, I. & Senthil, T. Mixed-valence insulators with neutral Fermi surfaces. *Nature Communications* **9**, 1766 (2018).
12  Shen, H. & Fu, L. Quantum Oscillation from In-Gap States and a Non-Hermitian Landau Level Problem. *Physical Review Letters* **121**, 026403 (2018).
13  Du, L. *et al.* Evidence for a topological excitonic insulator in InAs/GaSb bilayers. *Nature Communications* **8**, 1971 (2017).
14  Burg, G. W. *et al.* Strongly Enhanced Tunneling at Total Charge Neutrality in Double-Bilayer Graphene-WSe2 Heterostructures. *Physical Review Letters* **120**, 177702 (2018).
15  Wang, Z. *et al.* Evidence of high-temperature exciton condensation in two-dimensional atomic double layers. *Nature* **574**, 76-80 (2019).
16  Ma, L. *et al.* Strongly correlated excitonic insulator in atomic double layers. *Nature* **598**, 585-589 (2021).
17  Qi, R. *et al.* Thermodynamic behavior of correlated electron-hole fluids in van der Waals heterostructures. *Nature Communications* **14**, 8264 (2023).
18  Nguyen, P. X. *et al.* Perfect Coulomb drag in a dipolar excitonic insulator. *arXiv preprint arXiv:2309.14940* (2023).
19  Ruishi Qi, A. Y. J., Zuocheng Zhang, Jingxu Xie, Qixin Feng, Zheyu Lu, Ziyu Wang, Takashi Taniguchi, Kenji Watanabe, Sefaattin Tongay, Feng Wang. Perfect

| | |
|---|---|
| | Coulomb drag and exciton transport in an excitonic insulator. *arXiv:2309.15357* (2023). |
| 20 | Ruishi Qi, Q. L., Zuocheng Zhang, Sudi Chen, Jingxu Xie, Yunbo Ou, Zhiyuan Cui, David D. Dai, Andrew Y. Joe, Takashi Taniguchi, Kenji Watanabe, Sefaattin Tongay, Alex Zettl, Liang Fu, Feng Wang. Electrically controlled interlayer trion fluid in electron-hole bilayers. *arXiv:2312.03251* (2023). |
| 21 | Nguyen, P. X. *et al.* A degenerate trion liquid in atomic double layers. *arXiv preprint arXiv:2312.12571* (2023). |
| 22 | Zou, B., Zeng, Y., MacDonald, A. H. & Strashko, A. Electrical control of two-dimensional electron-hole fluids in the quantum Hall regime. *Physical Review B* **109**, 085416 (2024). |
| 23 | Shao, Y. & Dai, X. Quantum oscillations in an excitonic insulating electron-hole bilayer. *Physical Review B* **109**, 155107 (2024). |
| 24 | Han, Z., Li, T., Zhang, L., Sullivan, G. & Du, R.-R. Anomalous Conductance Oscillations in the Hybridization Gap of InAs/GaSb Quantum Wells. *Physical Review Letters* **123**, 126803 (2019). |
| 25 | Xiao, D., Liu, C.-X., Samarth, N. & Hu, L.-H. Anomalous Quantum Oscillations of Interacting Electron-Hole Gases in Inverted Type-II InAs/GaSb Quantum Wells. *Physical Review Letters* **122**, 186802 (2019). |
| 26 | Wang, R., Sedrakyan, T. A., Wang, B., Du, L. & Du, R.-R. Excitonic topological order in imbalanced electron–hole bilayers. *Nature* **619**, 57-62 (2023). |
| 27 | Wang, P. *et al.* Landau quantization and highly mobile fermions in an insulator. *Nature* **589**, 225-229 (2021). |
| 28 | Zhu, J., Li, T., Young, A. F., Shan, J. & Mak, K. F. Quantum oscillations in two-dimensional insulators induced by graphite gates. *Physical Review Letters* **127**, 247702 (2021). |
| 29 | Lee, P. A. Quantum oscillations in the activated conductivity in excitonic insulators: Possible application to monolayer WTe2. *Physical Review B* **103**, L041101 (2021). |
| 30 | He, W.-Y. & Lee, P. A. Quantum oscillation of thermally activated conductivity in a monolayer WTe2-like excitonic insulator. *Physical Review B* **104**, L041110 (2021). |
| 31 | Allocca, A. A. & Cooper, N. R. Quantum oscillations of magnetization in interaction-driven insulators. *Scipost Phys* **12** (2022). |
| 32 | Allocca, A. A. & Cooper, N. R. Fluctuation-dominated quantum oscillations in excitonic insulators. *Physical Review Research* **6**, 033199 (2024). |
| 33 | Zyuzin, V. A. de Haas--van Alphen effect and quantum oscillations as a function of temperature in correlated insulators. *Physical Review B* **109**, 235111 (2024). |
| 34 | Fogler, M. M., Butov, L. V. & Novoselov, K. S. High-temperature superfluidity with indirect excitons in van der Waals heterostructures. *Nature Communications* **5**, 4555 (2014). |
| 35 | Wu, F.-C., Xue, F. & MacDonald, A. H. Theory of two-dimensional spatially indirect equilibrium exciton condensates. *Physical Review B* **92**, 165121 (2015). |
| 36 | Xie, M. & MacDonald, A. H. Electrical Reservoirs for Bilayer Excitons. *Physical Review Letters* **121**, 067702 (2018). |
| 37 | Zeng, Y. & MacDonald, A. H. Electrically controlled two-dimensional electron-hole fluids. *Physical Review B* **102**, 085154 (2020). |
| 38 | Eisenstein, J. P. & MacDonald, A. H. Bose–Einstein condensation of excitons in bilayer electron systems. *Nature* **432**, 691-694 (2004). |
| 39 | Tiemann, L. *et al.* Exciton condensate at a total filling factor of one in Corbino two-dimensional electron bilayers. *Physical Review B* **77**, 033306 (2008). |
| 40 | Nandi, D., Finck, A. D. K., Eisenstein, J. P., Pfeiffer, L. N. & West, K. W. Exciton condensation and perfect Coulomb drag. *Nature* **488**, 481-484 (2012). |
| 41 | Liu, X., Watanabe, K., Taniguchi, T., Halperin, B. I. & Kim, P. Quantum Hall drag of exciton condensate in graphene. *Nature Physics* **13**, 746-750 (2017). |
| 42 | Li, J. I. A., Taniguchi, T., Watanabe, K., Hone, J. & Dean, C. R. Excitonic superfluid phase in double bilayer graphene. *Nature Physics* **13**, 751-755 (2017). |
| 43 | Stier, A. V. *et al.* Magnetooptics of Exciton Rydberg States in a Monolayer Semiconductor. *Physical Review Letters* **120**, 057405 (2018). |
| 44 | DinhDuy Vu, S. D. S. Excitonic phases in a spatially separated electron-hole ladder model. *arXiv:2305.16305* (2023). |
| 45 | Wang, L. *et al.* One-dimensional electrical contact to a two-dimensional material. *Science* **342**, 614-617 (2013). |
| 46 | Shen, P.-C. *et al.* Ultralow contact resistance between semimetal and monolayer semiconductors. *Nature* **593**, 211-217 (2021). |
| 47 | Ashoori, R. C. *et al.* Single-electron capacitance spectroscopy of discrete quantum levels. *Physical Review Letters* **68**, 3088-3091 (1992). |

# Figures

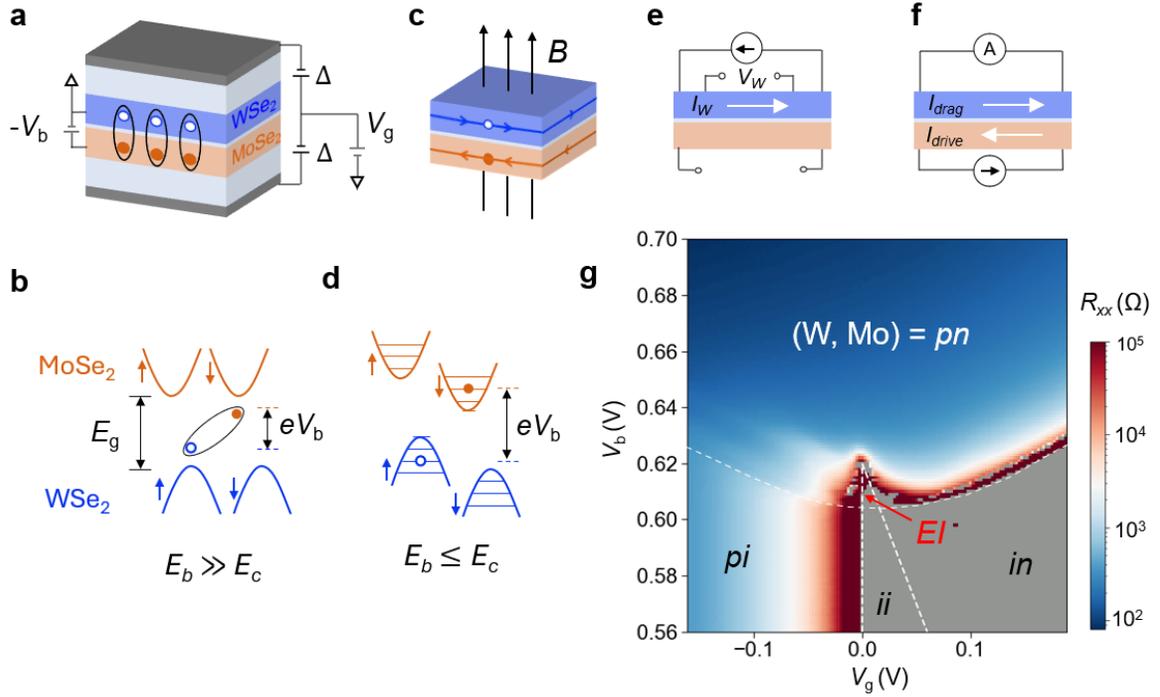

**Figure 1 | Experimental setup and electrostatics phase diagram. a,** Schematic of a dual-gated MoSe$_2$/WSe$_2$ double layer device. Interlayer bias voltage $V_b$ (with the W-layer grounded), $V_g \equiv \frac{V_{tg}+V_{bg}}{2}$ and $\Delta \equiv \frac{V_{tg}-V_{bg}}{2}$ control, respectively, the exciton chemical potential, net charge density and out-of-plane electric field in the Coulomb-coupled double layer ($V_{tg}$ and $V_{bg}$ denoting the top and bottom gate voltages, respectively). Interlayer excitons are stabilized at $B = 0$ T and low electron-hole pair densities. **b,** Band alignment of the double layer (not angle aligned) under $B = 0$ T or when the exciton binding energy $E_b$ far exceeds the cyclotron energy $E_c$. The conduction and valence bands are shown in orange and blue, respectively, the arrows denote the spin states, and the dashed lines denote the Fermi levels. Interlayer excitons spontaneously form when $E_b$ exceeds the effective charge gap $E_g - eV_b$. **c,** Same as **a** for high magnetic fields and high pair densities. Layer-decoupled QH states that support counter-propagating electron and hole chiral edge states emerge. **d,** Same as **b** for $E_b \leq E_c$. Horizontal solid lines label the electron and hole LLs. The states are spin-split by the Zeeman effect. **e,f,** Schematic of the open-circuit measurements of $R_{xx}$ and $R_{xy}$ in the W-layer (**e**) and the drag counterflow measurements (**f**). **g,** $R_{xx}$ as a function of $V_g$ and $V_b$ at $B = 0$ T and $T = 1.5$ K. The dashed lines separate the EI region and the different doping regimes of the Mo- and W-layers, *ii*, *in*, *pi* and *pn* (*i*, *p* and *n* denoting the intrinsic, hole-doped and electron-doped regimes, respectively). In the grey area, $R_{xx}$ diverges and cannot be reliably measured.

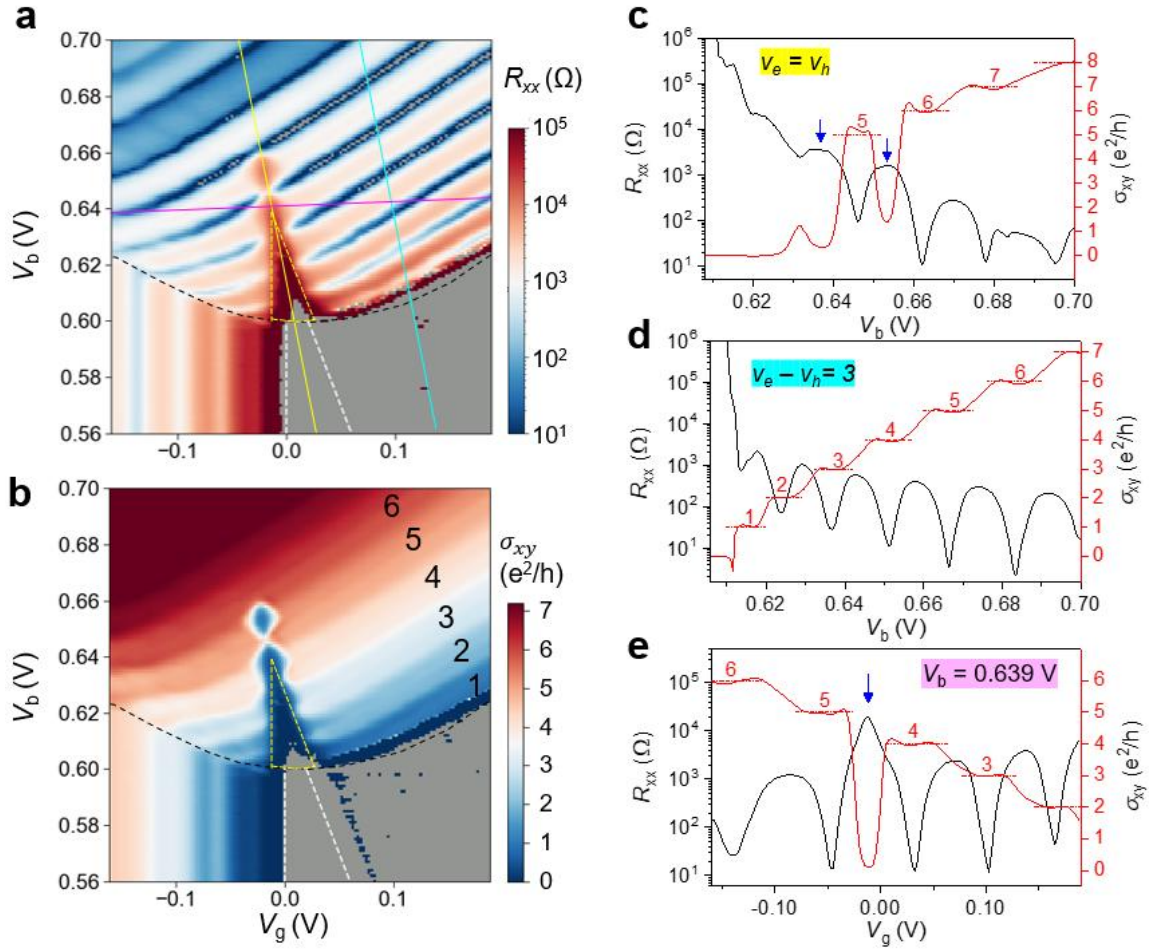

**Figure 2 | Quantum oscillations in the EI phase. a,b,** $R_{xx}$ (**a**) and $\sigma_{xy}$ (**b**) of the W-layer as a function of $V_g$ and $V_b$ at $B = 12$ T and $T = 1.5$ K. The triangle enclosed by the yellow dashed lines is the EI region. The $\sigma_{xy} = \frac{\nu_h e^2}{h}$ plateaus for $\nu_h = 1 - 6$ are labeled. In the grey area, $R_{xx}$ diverges and $R_{xx}$ and $\sigma_{xy}$ cannot be reliably measured. **c-e,** Linecuts of **a** and **b** along the solid yellow (**c**), cyan (**d**) and purple (**e**) lines in **a**. The QH states are characterized by $R_{xx}$ dips and quantized $\sigma_{xy}$ (horizontal dashed lines). Blue arrows denote the EI state (with $R_{xx}$ peaks and suppressed $\sigma_{xy}$) between two adjacent integer QH states.

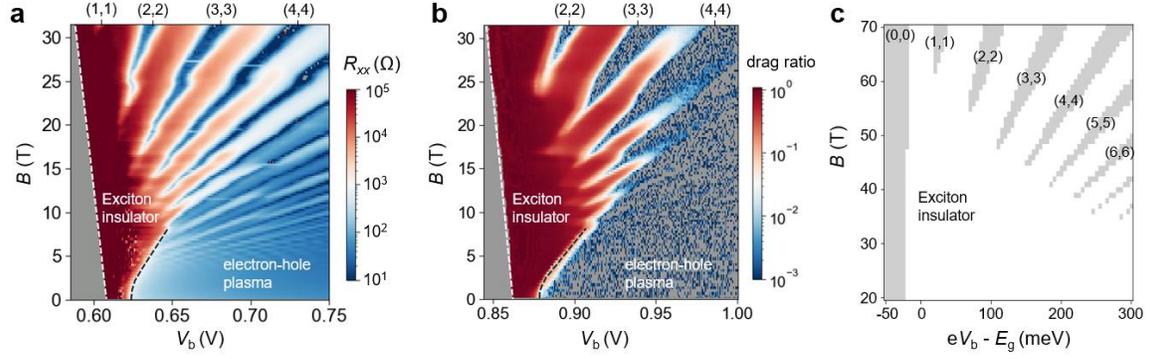

**Figure 3 | Magnetic-field phase diagram at charge neutrality. a,b,** $R_{xx}$ of the W-layer (device 1, **a**) and the drag ratio $\frac{I_{drag}}{I_{drive}}$ (device 3, **b**) as a function of $V_b$ and $B$ at charge neutrality (i.e. $\nu_e = \nu_h$) and $T = 0.3$ K. In the grey area, $R_{xx}$ diverges and cannot be reliably measured. The white dashed line denotes the band edge, and the black dashed line separates the EI and electron-hole plasma phases. The EI phase, which has diverging $R_{xx}$ and large $\frac{I_{drag}}{I_{drive}}$, expands with magnetic field. The fully filled LLs that protrude into the EI phase, $(\nu_e, \nu_h) = (1,1), (2,2), (3,3), (4,4)$, which have vanishing $R_{xx}$ and $\frac{I_{drag}}{I_{drive}}$, can be identified at high fields. **c,** Mean-field phase diagram in exciton chemical potential ($eV_b - E_g$) and $B$ at charge neutrality (see Methods). White region: the EI state; Grey: layer-decoupled QH states.

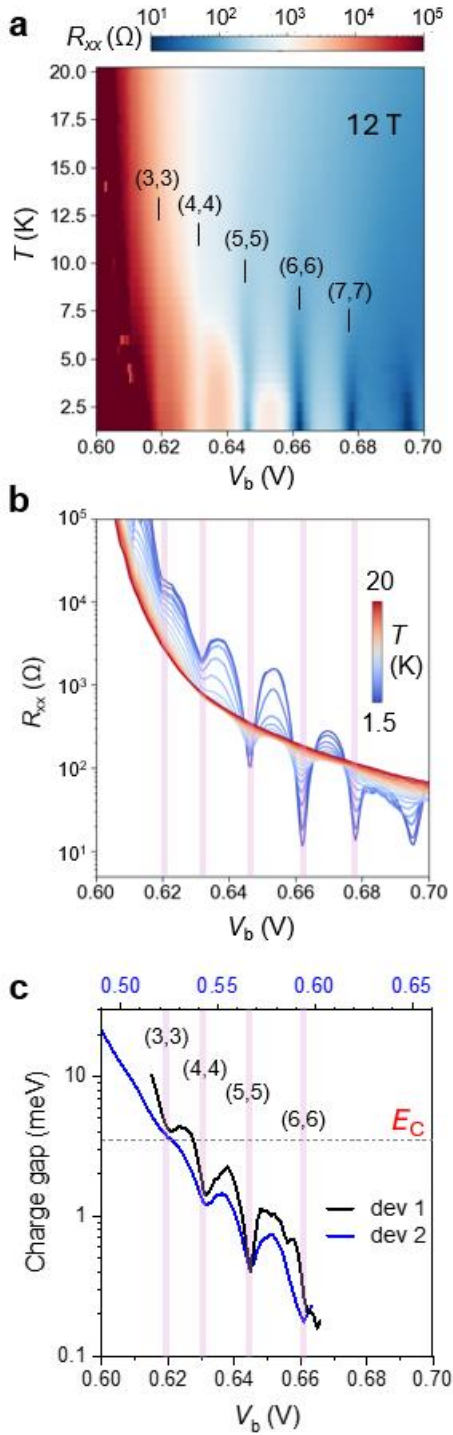

**Figure 4 | Oscillating charge gap (B = 12 T). a,** $R_{xx}$ of the W-layer as a function of $V_b$ and $T$ at charge neutrality. **b,** Linecuts of **a** at representative temperatures. The EI state (with increasing $R_{xx}$ as $T$ decreases) transitions to the QH states (with decreasing $R_{xx}$ as $T$ decreases) as $V_b$ increases. **c,** Thermodynamic gap size for device 1 and 2 as a function of $V_b$. The horizontal dashed line marks the cyclotron energy. There are oscillations in the gap size on top of a monotonically decreasing background. The fully filled LLs are labeled in **a** and denoted by purple lines in **b,c**.

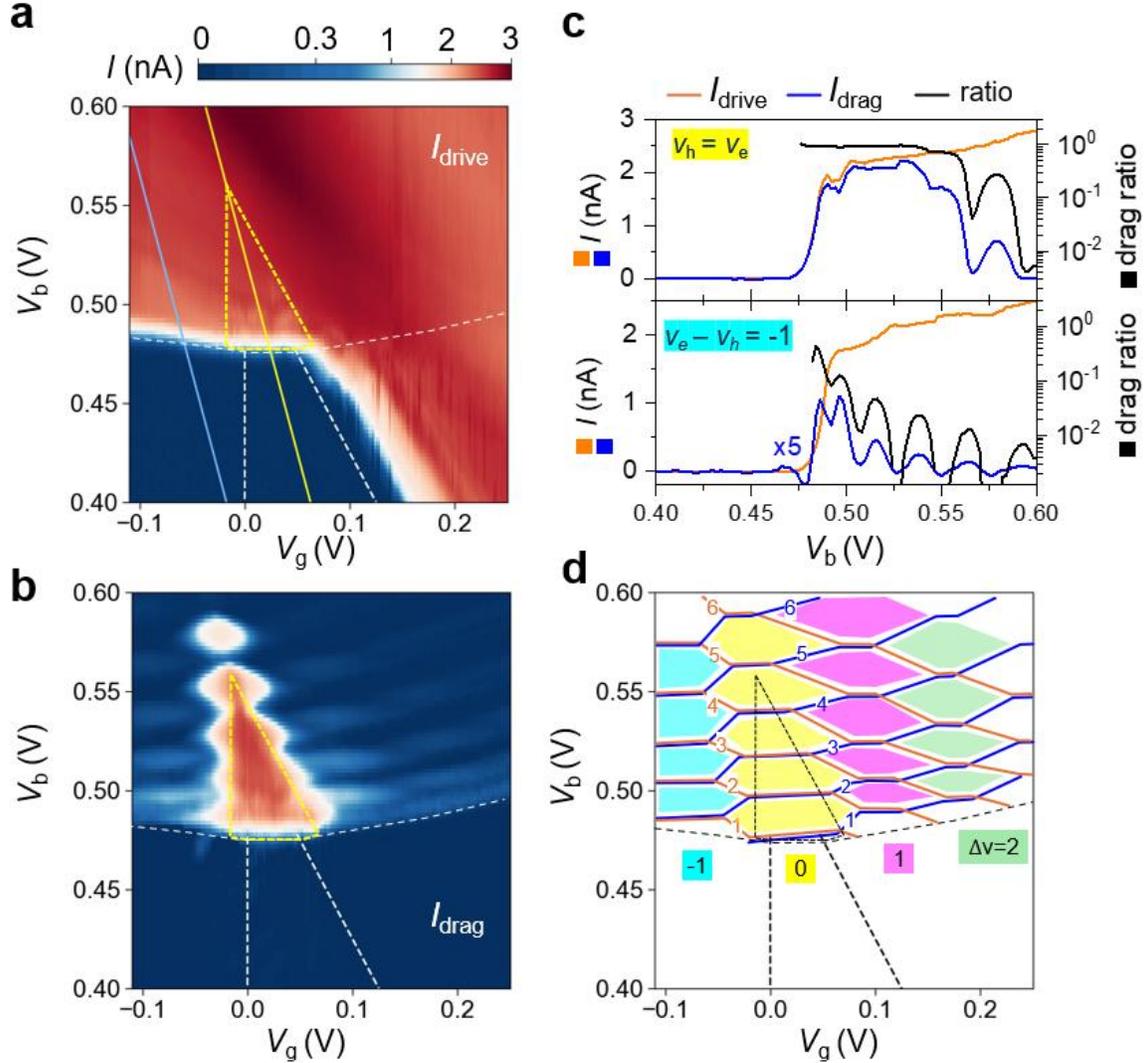

**Figure 5 | Coulomb drag and finite angular momentum excitons. a,b,** Drive current ($I_{drive}$) in the Mo-layer (**a**) and drag current ($I_{drag}$) in the W-layer (**b**) as a function of $V_g$ and $V_b$ at $B = 12$ T and $T = 1.5$ K. Near perfect Coulomb drag is observed inside the EI region bound by the yellow dashed lines. Substantial Coulomb drag is also observed outside the EI region with partially filled LLs, indicating the formation of finite angular momentum excitons. **c,** Linecuts of **a,b** and drag ratio $\frac{I_{drag}}{I_{drive}}$ along the solid yellow line (upper panel with $\nu_e = \nu_h$) and cyan line (lower panel with $\nu_e - \nu_h = -1$) in **a**. An enhanced drag ratio indicates stronger interlayer excitonic interactions. The layer-decoupled QH states manifest significant drops in the drag ratio. **d,** Phase diagram for the QH states in each layer (orange and blue lines for the Mo- and W-layer, respectively) constructed from **b** (see main text). The regions with $\nu_e - \nu_h = -1, 0, 1$ and $2$ are shaded in cyan, yellow, purple and green, respectively. The EI state (triangle bound by the dashed lines) dominates the $\nu_e = \nu_h$ region.

**Extended Data Figures**

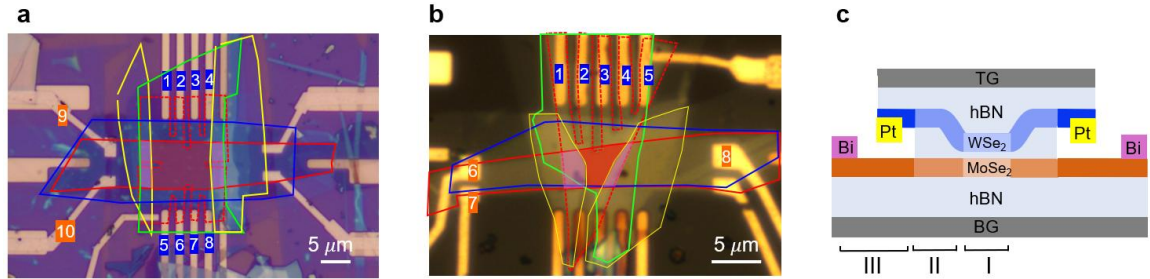

**Extended Data Figure 1 | Device images and schematic. a,b,** Optical micrographs for device 1 (**a**) and 2 (**b**). The top gate (red dashed line), bottom gate (red line), W-layer (green line), Mo-layer (blue line) and the exciton contact hBN layers (yellow lines) are outlined. The red- and pink-shaded areas denote the device channel and exciton contact regions, respectively. The contact electrodes to the Mo- and W-layers are numbered in orange and blue boxes, respectively. **c**, Schematic of the device cross section showing the doping profile. Region I, II and III are the channel, the exciton contact and the metal-semiconductor contacts, ordered in increasing doping concentration represented by the lighter to darker orange/blue color (orange for electron, blue for hole doping). TG, BG, Pt and Bi denote top gate, bottom gate, Platinum electrode and Bismuth electrode, respectively.

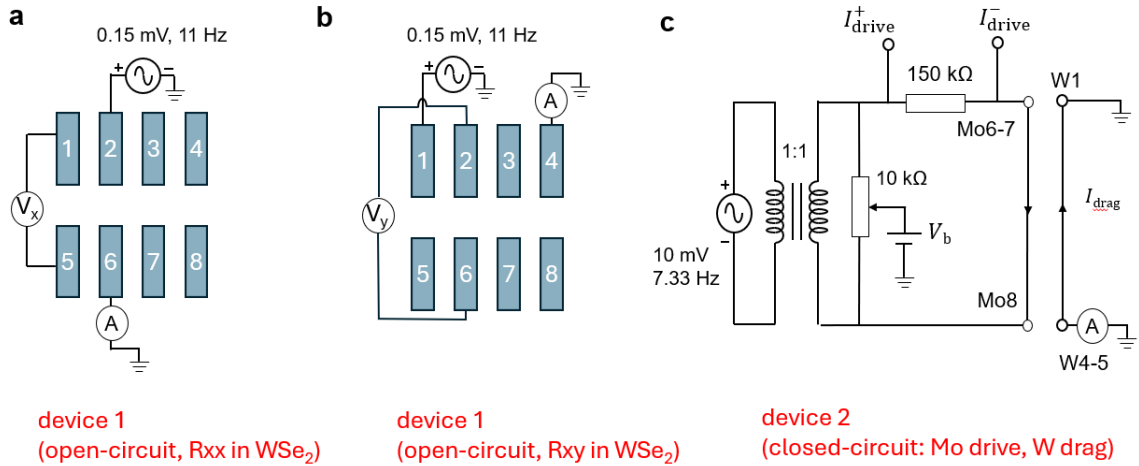

**Extended Data Figure 2 | Measurement circuit diagrams. a,b,** Open-circuit measurement diagrams for $R_{xx}$ (**a**) and $R_{xy}$ (**b**) in device 1. The Mo-layer is in open-circuit; only pin 9 is connected to apply a DC interlayer bias $V_b$ (not shown). (See pin numbers in Extended Data Fig. 1). We only show the pin connections to measure the $R_{xx}$ (**a**) and $R_{xy}$ (**b**) in the W-layer. **c,** Closed-circuit measurement diagram for the drag counterflow studies in device 2. An AC electron current is biased in the Mo-layer through a 1:1 transformer. $V_b$ is applied (through the midpoint of a 10 kΩ potentiometer) to both the source (shorted pins 6 and 7) and drain (pin 8) in the Mo-layer. The potentiometer position is tuned to minimize the AC interlayer coupling. In the W-layer, pin 1 is grounded and the hole drag current is measured by an ammeter connected to pins 4 and 5 (shorted).

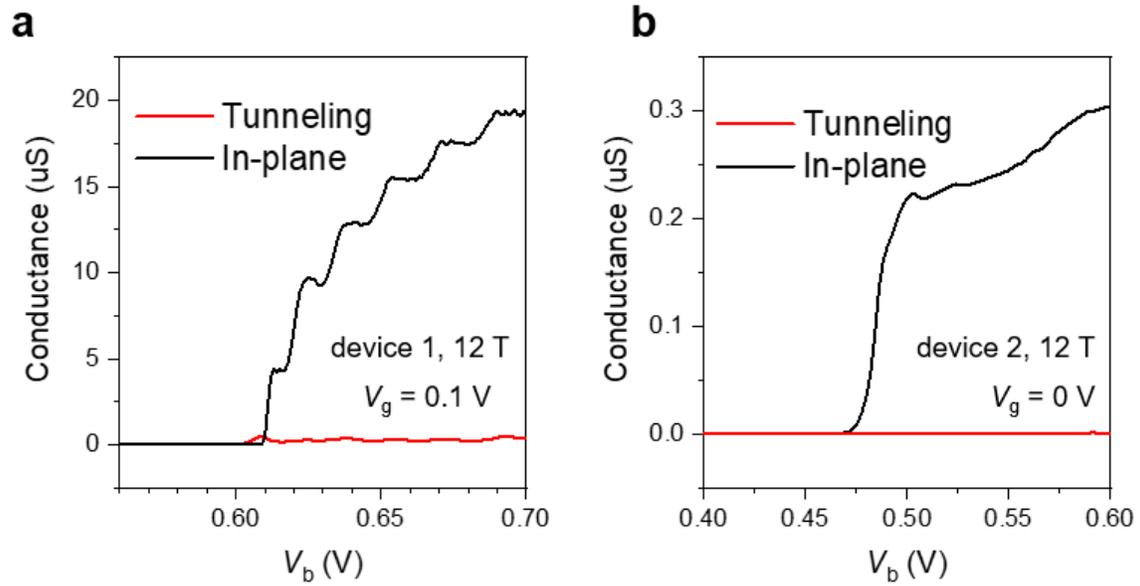

**Extended Data Figure 3 | Negligible interlayer tunneling conductance. a,b,** Interlayer tunneling conductance (red) and two-terminal in-plane conductance of the Mo-layer (black) as a function of $V_b$ for device 1 at $V_g = 0.1$ V (**a**) and device 2 at $V_g = 0$ V (**b**). In the relevant *pn* and EI regions (i.e. when the Mo-layer is turned on electrically), both devices show tunneling conductance about 1-2 orders of magnitude smaller than the in-plane conductance.

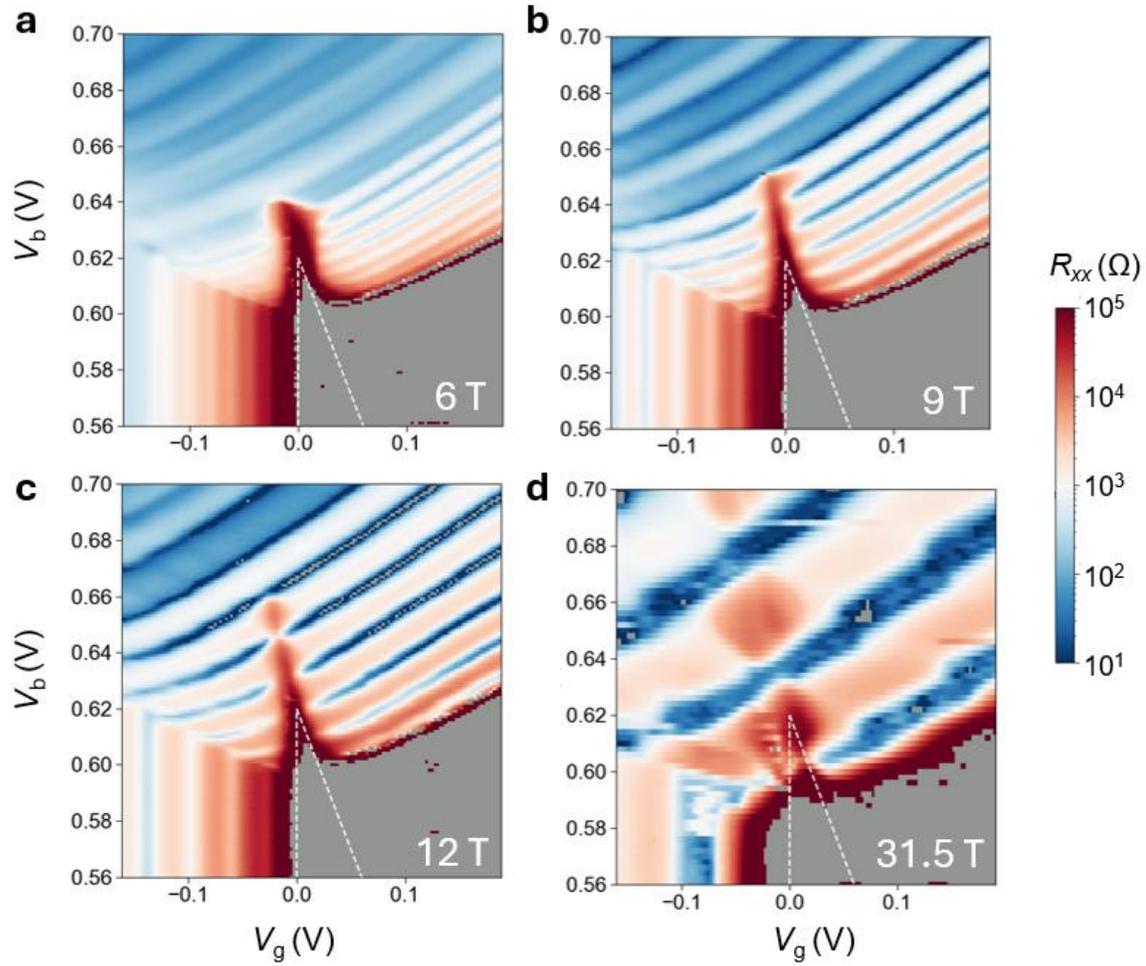

**Extended Data Figure 4 | Magnetic field evolution of $R_{xx}$. a-d,** $R_{xx}$ of the W-layer as a function of $V_g$ and $V_b$ at $B = 6$ T (**a**), 9 T (**b**), 12 T (**c**) and 31.5 T (**d**). The dashed lines locate the band edges under zero magnetic field. $R_{xx}$ diverges and cannot be reliably measured in the grey area. The dips in $R_{xx}$ reflect the fully filled LLs. The voltage spacing between consecutive dips increases with $B$ as the LL degeneracy $\Delta n_{LL} = \frac{eB}{h}$ increases. At $B = 31.5$ T, the lowest LL penetrates through the EI triangular region.

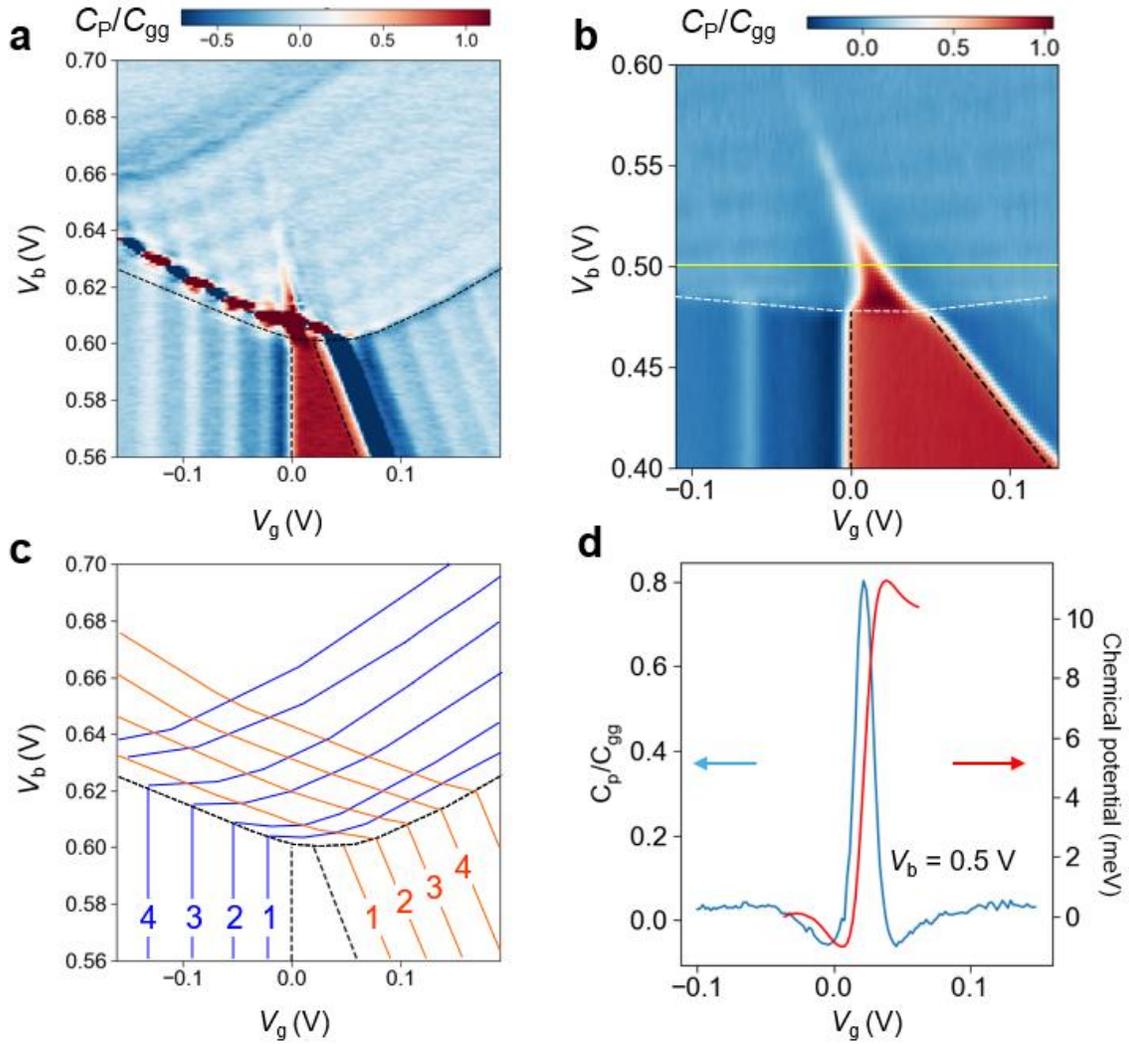

**Extended Data Figure 5 | Penetration capacitance measurements. a,b,** The normalized penetration capacitance $C_P/C_{gg}$ as a function of $V_g$ and $V_b$ at $B = 12$ T and $T = 1.5$ K for device 1 (**a**) and 2 (**b**). LLs from both the W- and Mo-layers can be observed, especially in device 1. The electronic incompressibility of the EI oscillates as $V_b$ increases at charge-neutrality. **c,** The extracted integer-filled LLs for the Mo- (orange) and W-layer (blue) from the data in **a**. **d,** $C_P/C_{gg}$ (blue) and the extracted electron chemical potential (red) as a function of $V_g$ along the yellow solid line in **b**. The chemical potential is obtained by integrating $C_P/C_{gg}$ with respect to $V_g$ (see Methods). The thermodynamic gap size is equal to the jump in the chemical potential for the incompressible state.

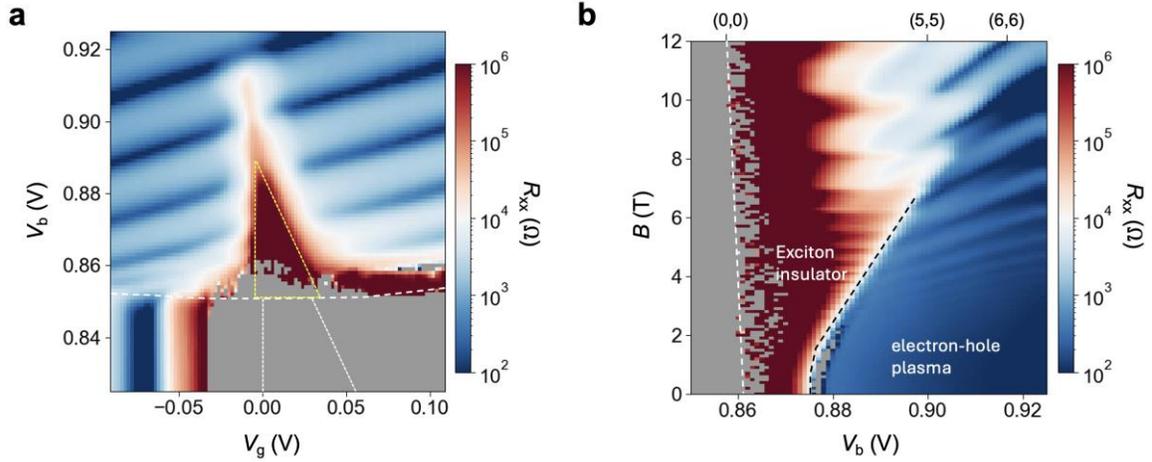

**Extended Data Figure 6 | Reproducible data in device 3. a,** $R_{xx}$ of the W-layer as a function of $V_g$ and $V_b$ at $B = 12$ T and $T = 1.5$ K. The triangular region enclosed by the yellow dashed lines is the EI region. **b,** $R_{xx}$ as a function of $V_b$ and $B$ at charge neutrality (i.e. $v_e = v_h$) and $T = 1.5$ K. The EI boundary (black dashed line) expands with magnetic field. The fully filled LLs that protrude into the EI phase at high fields are labelled as $(v_e, v_h) = (5,5), (6,6)$; the band edge is labeled as $(v_e, v_h) = (0,0)$. The grey-shaded areas denote diverging $R_{xx}$ that cannot be reliably measured.